# First Beam Characterization by Means of Emission Spectroscopy in the NIO1 Experiment

M. Barbisan[1, a)], B. Zaniol[1], M. Cavenago[2], G. Serianni[1] and R. Pasqualotto[1]

[1] *Consorzio RFX (CNR, ENEA, INFN, University of Padova, Acciaierie Venete SpA), C.so Stati Uniti 4 – 35127, Padova (Italy).*
[2] *INFN-LNL, v.le dell'Università 2, I-35020, Legnaro (PD) Italy.*

[a)]Corresponding author: M. Barbisan, e-mail address: barbisan@igi.cnr.it

**Abstract.** The NIO1 experiment hosts a flexible RF $H^-$ ion source, developed by INFN-LNL and Consorzio RFX to improve the present concepts for the production and acceleration of negative ions. The source is also used to benchmark the instrumentation dedicated to the ITER neutral beam test facility. Many diagnostics are installed in NIO1 to characterize the source and the extracted negative ion beam. Among them, Beam Emission Spectroscopy (BES) has been used in NIO1 to measure the divergence and the uniformity of the beam, together with the fraction of beam ions which was neutralized inside the acceleration system. The diagnostic method is based on the analysis of the Doppler shifted $H_\alpha$ photons emitted by the fast beam particles and collected along a line of sight. The article presents the experimental setup and the analysis algorithms of the BES diagnostic, together with a discussion of the first measurements and of their correlation with the operational parameters.

## INTRODUCTION

In the framework of the research on Neutral Beam Injectors (NBIs) for fusion reactors, Consorzio RFX and INFN-LNL are exploiting the NIO1 (Negative Ion Optimization 1) negative ion source [1]. Target of the experiment is to optimize the production and acceleration of negative ions ($H^-$) in the NBIs of future fusion reactors. NIO1 is also supporting the construction of the ITER neutral beam test facility, with tests on part of the diagnostic instrumentation dedicated to it. NIO1 consists of an Inductively Coupled Plasma (ICP) ion source, cylindrically shaped (10 cm diameter x 21 cm length), in which a hydrogen plasma is created thanks to the radiofrequency (RF) generated by a 2 MHz, up to 2.5 kW transmitter and irradiated by a solenoid. The negative ions produced in the plasma are extracted and accelerated by a set of grids: Plasma Grid (facing the source), Extraction Grid, Post Acceleration Grid (at ground voltage) and Repeller. The grids feature a square lattice of 3x3 apertures, through which as many beamlets are produced. The horizontal and vertical spacing between adjacent apertures is 14 mm, while the diameter of the final apertures on the repeller is 8.8 mm. With the future evaporation in the source of Cs to enhance $H^-$ surface production, NIO1 is expected to produce a maximum beam current of 130 mA, composed by $H^-$ ions accelerated at maximum 60 keV, in continuous (>1000 s) operation.

Source plasma and extracted negative ion beam are characterized by several diagnostics to study the effects of modifications of the experiment operative parameters or of hardware upgrades of NIO1 [2]. Beam diagnostics include a 1D and a 2D Charge-Coupled Device (CCD) camera, Beam Emission Spectroscopy (BES), and a Carbon Fiber Composite (CFC) tile, positioned about 40 cm downstream along the acceleration system. The CFC tile blocks the beam particles and allows a measurement of the 2D beam power distribution by means of infrared (IR) thermography. BES is exploited in NIO1 to measure direction, divergence and intensity of the beam, together with the fraction of negative ions which have been neutralized inside the acceleration system (stripping losses). BES diagnostic design was presented and discussed in [3]. This paper describes the installed instrumentation and setup of the BES diagnostic, together with the data analysis method and a preliminary discussion of the first data.

# THE EXPERIMENTAL SETUP FOR BES IN NIO1

The $H_\alpha$ radiation emitted by the beam is collected by an optic head, whose Line of Sight (LoS) is oriented about 60° in the horizontal plane with respect to the beam axis (Fig. 1(a)), the orientation closest to the design value of 75° [3], constrained by the position of the CFC tile inside the vacuum chamber. The optic head hosts a lens with focal length 50 mm and clear aperture diameter 6 mm, focusing the collected light on a 400 μm core diameter silica-silica fiber with 0.22 NA. The light is conveyed though the fiber to an *Isoplane SCT 320* spectrometer [4], with entrance slit width set at 50 μm and mounting a 2000 gr/mm grating. The spectrometer is coupled to a *PIXIS 2K/BUV* camera, equipped with a back illuminated CCD of 2048x512 pixels, 13.5 μm x 13.5 μm large [4]. The plate factor of the system given by spectrometer and camera is 12.9 pm/pixel (at 656 nm). The instrument function of the system, measured by studying the imaging of a spectral line with negligible intrinsic width (namely the 508.5 nm line of Cd), has a FWHM width of 5.05±0.01 pixels. In the present experimental campaigns, despite the total acceleration voltage below 10 kV and the accelerated beam current usually below 5 mA, an exposure time of 8 s for the CCD was adequate to collect BES spectra with a SNR suitable for the data analysis.

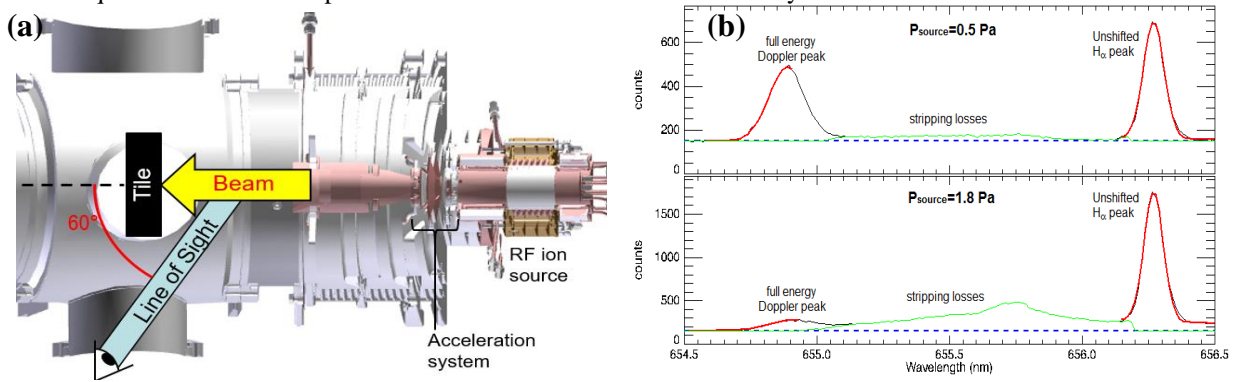

**FIGURE 1.** (a) 3D horizontal section of the NIO1 experiment and schematic representation of beam and BES LoS directions. (b) Example spectra acquired by BES diagnostic; the black curves indicate the acquired spectra, the blue dashed lines indicate the linear estimations of the background signals, the red curves indicate the Gaussians fitted to the unshifted $H_\alpha$ and full energy Doppler peaks. The green curve indicates the part of the Doppler emissions attributed to stripping losses. The top and bottom spectra were acquired at the same NIO1' operational parameters, except source pressure (0.5 Pa and 1.8 Pa, respectively).

# ANALYSIS OF BES SPECTRA AND RESULTS

Fig. 1(b) shows two example BES spectra, acquired in NIO1 under the same experimental conditions (RF power 1.2 kW, extraction voltage 1.3 kV, total acceleration voltage 8 kV) except for the source pressure $P_{source}$, set at 0.5 Pa in the top spectrum and at 1.8 Pa in the bottom spectrum, respectively. The unshifted $H_\alpha$ line at 656.28 nm is emitted by the slow H atoms produced by the dissociation of $H_2$ molecules, as effect of the collision between these molecules and the beam particles. The same collisions can lead to the neutralization and to the excitation of fast beam particles, which leads to emissions with measurable Doppler shift: the Doppler peak given by fully accelerated beam particles and the broad emission of stripping losses, at intermediate values of Doppler shift. As expected, the emissions of stripping losses represent a higher fraction of Doppler shifted emissions in the case at higher $P_{source}$ in Fig. 1(b), since the gas density (and then the neutralization rate of $H^-$) in the acceleration system grows with $P_{source}$.

BES spectra analysis firstly consists in subtracting the background level, fitted with a line (blue dashed line in Fig.1(b)). The full energy Doppler peak and the unshifted $H_\alpha$ line are fitted with Gaussian curves to retrieve central wavelength and width of the 2 peaks (red lines in Fig.1(b)). In the case of the full energy Doppler peak, the wavelength interval selected for the fit does not cover part of the low energy side (i.e. the one at higher wavelengths) of the peak, since the overlapping of the stripping losses emission may spoil the fit results. At last, the emissions of stripping losses (green curve in Fig. 1(b)) are estimated after subtraction of the two fitted Gaussian curves.

The information which can be retrieved from the BES spectra is as follows: from the wavelength difference between the full energy Doppler peak and the unshifted $H_\alpha$ peak the average direction of beamlets can be measured [3]. The integral of the full energy Doppler peak is an indirect measurement of beam density. The ratio between the integrals of stripping losses and the full energy Doppler peak can be used to study the variations in the fraction of stripping losses. Most importantly, from the width of the full energy Doppler peak it is possible to estimate the beam divergence [3,5]. To obtain this estimate, the broadening of the peak due to beam divergence must be separated from other broadening factors: the intrinsic width of the line (0.0132 nm as sigma width), the instrument function of the spectrometer (5.05±0.01 pixels FWHM), the ripple of the high voltage power supplies (0.2%±0.02%) and the aperture angle of the collection optics (8.67 mrad ±10%).

Examples of BES results obtained during NIO1 experimental campaign are shown in Fig.2; plot (a) shows the ratio between the integrals of the stripping loss emissions and of the full energy Doppler peak, for a scan in source pressure (from 0.5 Pa to 1.8 Pa) at 1.2 kW RF power, 1.3 kV extraction voltage and 8 kV total acceleration voltage. The points at extreme $P_{source}$ values were obtained from the spectra shown in Fig. 1(b). The trend of the data shown in Fig. 2(a) indicates, as expected, that the fraction of stripping losses steadily increases with $P_{source}$, as a consequence of the increased gas density in the acceleration system. Fig. 2(b) instead shows BES data belonging to a scan in EG voltage (from 700 V to 1.4 kV), at fixed RF power (1.2 kW) and total acceleration voltage (8 kV), and at $P_{source}$=1 Pa. The beam divergence (void markers) and the full energy Doppler peak integral (solid markers) are plotted as function of the beam perveance (P) in the PG-EG gap, normalized to the Child-Langmuir perveance ($P_0$). The plot shows that with lower $P/P_0$ (i.e. higher EG voltage) a higher beam intensity is achieved. Moreover, the BES estimates of beam divergence have absolute values around 40 mrad, increasing with $P/P_0$. According to the model [5] used to estimate beam divergence and to the uncertainties estimated for its input parameters, the divergence measurements have relative errors around 5%.

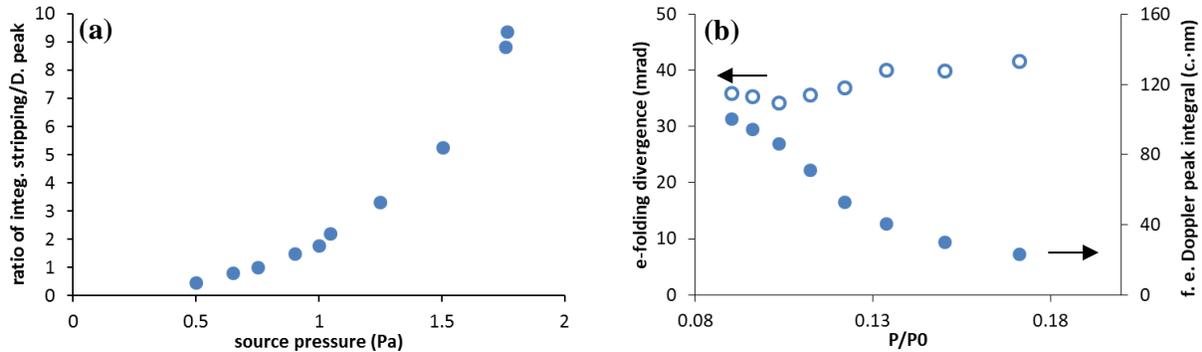

**FIGURE 2.** Plot (a): Ratio of the integrals of stripping losses and full energy Doppler peak, plotted as function of the source pressure. The data belong to a scan in $P_{source}$. Plot(b): e-folding divergence (void markers) and full energy Doppler peak integral (solid markers), plotted against $P/P_0$ in the extraction gap. The data belong to a scan in extraction voltage, at $P_{source}$=1 Pa.

## CONCLUSIONS

The BES diagnostic has been successfully installed in the NIO1 experiment and is routinely working during NIO1 experimental campaigns. Beam divergence, beam horizontal aiming and relative variations of beam intensity and stripping fraction can be measured. As preliminary check, known relations between the BES results and the operational parameters have been successfully verified. In future, the results of BES diagnostic will be validated by comparing them with the output of the other beam diagnostics and with numeric simulations of the beam.

## ACKNOWLEDGMENTS

This work has been carried out within the framework of the EUROfusion Consortium and has received funding from the Euratom research and training programme 2014-2018 under grant agreement No 633053. The views and opinions expressed herein do not necessarily reflect those of the European Commission.